\documentclass[submission]{eptcs}
\providecommand{\event}{SSV 2012} 
\usepackage{breakurl}             
\usepackage{isabelle}
\usepackage{isabellesym}
\usepackage{amssymb}
\usepackage{amsmath}
\usepackage{mathpartir}
\usepackage{afterpage}
\usepackage[bottom]{footmisc}

\newcommand{\isafun}[1]{{\sf #1}}

\isabellestyle{it}

\title{Automatic Function Annotations for Hoare Logic}
\author{Daniel Matichuk
\institute{NICTA\\ Sydney, Australia}
\email{daniel.matichuk@nicta.com.au}
}
\def\titlerunning{Annotations}
\def\authorrunning{D. Matichuk}
\begin{document}
\maketitle

\begin{abstract}
In systems verification we are often concerned
with multiple, inter-dependent properties that a program must
satisfy. To prove that a program satisfies a given property,
the correctness of intermediate states of the program must be
characterized. However, this intermediate reasoning is not
always phrased such that it can be easily re-used in the
proofs of subsequent properties. We introduce a function
annotation logic that extends Hoare logic in two important ways:
(1) when proving that a function satisfies a Hoare triple, intermediate
reasoning is automatically stored as function annotations, and (2) these function
annotations can be exploited in future Hoare logic proofs.
This reduces duplication of reasoning between the proofs of different properties, whilst
serving as a drop-in replacement for traditional Hoare logic to
avoid the costly process of proof refactoring. We explain
how this was implemented in Isabelle/HOL and applied to an
experimental branch of the seL4 microkernel to significantly reduce
the size and complexity of existing proofs.
\end{abstract}

\section{Introduction}
\label{s:intro}

It is not always apparent
what properties need to be proven when formally verifying real software. Clearly
a program should maintain the consistency of its data structures, but it may
be less obvious to show that, for example, a scheduling policy is fair.
As a result, a verified system will inevitably be reasoned about multiple
times while considering different properties. This will
involve characterizing the correctness of intermediate states of
a program with 
respect to each of these properties. Likely these properties will have
some inter-dependencies, e.g. we can only reason that a scheduling
policy is fair if the scheduler queues are valid. In these cases,
the proofs of these properties will depend on results previously shown.
This inter-dependence of reasoning demonstrates a necessity to
structure theories and lemmas in such a way that maximizes their re-use.

The seL4 microkernel is to our knowledge the most extensive
code-level verification ever performed of a general-purpose
operating system kernel. The C implementation is shown to formally
\emph{refine} an abstract specification of its behaviour~\cite{Klein_EHACDEEKNSTW_09}, which is proven to preserve a set of kernel invariants. These
 invariants describe the correctness of all kernel structures,
  saying, for example, that distinct objects in memory do not
  overlap. In the subsequent proofs of security properties, namely
integrity and authority confinement~\cite{Sewell_WGMAK}, the existing lemmas shown
during the invariant proofs were heavily used.

The proofs of these properties were done using a monadic
variant of Hoare logic~\cite{Cock_KS_08}
, which is used to 
reason about pre and post conditions of functions. When an invariant is temporarily
violated, such as during the creation of new kernel objects, care must
be taken to describe this intermediate state, so it can be reasoned
that further operations re-establish this invariant. In 
traditional Hoare logic,
 correctness of this intermediate state cannot easily be expressed. 
If we wish to re-establish this temporary invariant violation in subsequent proofs,
we will be forced to perform the same reasoning required in the original proof.
 With powerful vcgs (verification condition generators)
and automated reasoning inside a theorem prover, the impact
of this duplicated reasoning can be minimal. However, for
a sufficiently complex function with corresponding complex
intermediate states, significant manual effort is required to
prove any given property. Re-establishing the correctness of these
intermediate states can then result in significant portions
of duplicated proof, which become difficult and costly to maintain
as a project evolves.

In this paper, we present a function annotation logic, which
allows the correctness of these intermediate states to be shown
as function annotations. Properties proven
about these intermediate states can then be used in future proofs
without any redundant reasoning. These annotations
do not need to be explicitly defined, but can be generated
as a consequence of an existing Hoare logic proof.

Function annotations are not novel, in~\cite{Hoare:1983:ABC:357980.358001} Hoare advocates proving the correctness of assertions so they
may be assumed in further proofs.
Traditionally, however, pre and post conditions are favoured over
manually annotating entire functions.
In existing verification frameworks, such as VCC~\cite{Cohen_DHLMSST_09}, annotations are
defined alongside the code and a function receives a
single set of annotations. The primary contribution
of this paper is the automatic extraction of
annotations from existing Hoare logic proofs.
Additionally, multiple sets of
annotations can be defined for orthogonal properties about a function.

\begin{isabellebody}%
\def\isabellecontext{Annotations}%
\isadelimtheory
\endisadelimtheory
\isatagtheory
\endisatagtheory
{\isafoldtheory}%
\isadelimtheory
\endisadelimtheory
\begin{isamarkuptext}%
\section{An example specification}
\label{s:example}

\begin{figure}[tb]
\begin{minipage}[t]{0.5\textwidth}\small
\begin{isabelle}%
\isafun{new{\isaliteral{5F}{\isacharunderscore}}tcb}\ p\ {\isaliteral{5C3C65717569763E}{\isasymequiv}}\ \isafun{do}\isanewline
\ \ \ \ i\ {\isaliteral{5C3C6C6566746172726F773E}{\isasymleftarrow}}\ \isafun{alloc}{\isaliteral{3B}{\isacharsemicolon}}\isanewline
\isaindent{\ \ \ \ }tcb\ {\isaliteral{5C3C6C6566746172726F773E}{\isasymleftarrow}}\ \isafun{create{\isaliteral{5F}{\isacharunderscore}}tcb}\ p{\isaliteral{3B}{\isacharsemicolon}}\isanewline
\isaindent{\ \ \ \ }\isafun{init{\isaliteral{5F}{\isacharunderscore}}tcb}\ tcb\ i{\isaliteral{3B}{\isacharsemicolon}}\isanewline
\isaindent{\ \ \ \ }\isafun{enqueue{\isaliteral{5F}{\isacharunderscore}}tcb}\ i\ p{\isaliteral{3B}{\isacharsemicolon}}\isanewline
\ \ \ \ \isafun{return}\ i\isanewline
\isafun{od}%
\end{isabelle}
\begin{isabelle}%
\isafun{alloc}\ {\isaliteral{5C3C65717569763E}{\isasymequiv}}\ \isafun{do}\isanewline
\ \ \ \ ids\ {\isaliteral{5C3C6C6566746172726F773E}{\isasymleftarrow}}\ \isafun{gets}\ \isafun{ids}{\isaliteral{3B}{\isacharsemicolon}}\isanewline
\isaindent{\ \ \ \ }i\ {\isaliteral{5C3C6C6566746172726F773E}{\isasymleftarrow}}\ \isafun{select}\ ids{\isaliteral{3B}{\isacharsemicolon}}\isanewline
\isaindent{\ \ \ \ }\isafun{set{\isaliteral{5F}{\isacharunderscore}}ids}\ {\isaliteral{28}{\isacharparenleft}}ids\ {\isaliteral{2D}{\isacharminus}}\ {\isaliteral{7B}{\isacharbraceleft}}i{\isaliteral{7D}{\isacharbraceright}}{\isaliteral{29}{\isacharparenright}}{\isaliteral{3B}{\isacharsemicolon}}\isanewline
\ \ \ \ \isafun{return}\ i\isanewline
\isafun{od}%
\end{isabelle}
\end{minipage}
\begin{minipage}[t]{0.5\textwidth}\small
\begin{isabelle}%
\isafun{create{\isaliteral{5F}{\isacharunderscore}}tcb}\ p\ {\isaliteral{5C3C65717569763E}{\isasymequiv}}\ \isafun{do}\isanewline
\ \ \ \ tcb\ {\isaliteral{5C3C6C6566746172726F773E}{\isasymleftarrow}}\ \isafun{return}\ \isafun{empty{\isaliteral{5F}{\isacharunderscore}}tcb}{\isaliteral{3B}{\isacharsemicolon}}\isanewline
\ \ \ \ \isafun{return}\ {\isaliteral{28}{\isacharparenleft}}tcb{\isaliteral{5C3C6C706172723E}{\isasymlparr}}priority\ {\isaliteral{3A}{\isacharcolon}}{\isaliteral{3D}{\isacharequal}}\ p{\isaliteral{5C3C72706172723E}{\isasymrparr}}{\isaliteral{29}{\isacharparenright}}\isanewline
\isafun{od}%
\end{isabelle}
\begin{isabelle}%
\isafun{init{\isaliteral{5F}{\isacharunderscore}}tcb}\ tcb\ i\ {\isaliteral{5C3C65717569763E}{\isasymequiv}}\ \isafun{do}\isanewline
\ \ \ \ tcbs\ {\isaliteral{5C3C6C6566746172726F773E}{\isasymleftarrow}}\ \isafun{gets}\ \isafun{tcbs}{\isaliteral{3B}{\isacharsemicolon}}\isanewline
\ \ \ \ \isafun{set{\isaliteral{5F}{\isacharunderscore}}tcbs}\ {\isaliteral{28}{\isacharparenleft}}tcbs{\isaliteral{28}{\isacharparenleft}}i\ {\isaliteral{5C3C6D617073746F3E}{\isasymmapsto}}\ tcb{\isaliteral{29}{\isacharparenright}}{\isaliteral{29}{\isacharparenright}}\isanewline
\isafun{od}%
\end{isabelle}
\begin{isabelle}%
\isafun{enqueue{\isaliteral{5F}{\isacharunderscore}}tcb}\ i\ p\ {\isaliteral{5C3C65717569763E}{\isasymequiv}}\ \isafun{do}\isanewline
\ \ \ \ qs\ {\isaliteral{5C3C6C6566746172726F773E}{\isasymleftarrow}}\ \isafun{gets}\ \isafun{queues}{\isaliteral{3B}{\isacharsemicolon}}\isanewline
\isaindent{\ \ \ \ }q\ {\isaliteral{5C3C6C6566746172726F773E}{\isasymleftarrow}}\ \isafun{return}\ {\isaliteral{28}{\isacharparenleft}}qs\ p{\isaliteral{29}{\isacharparenright}}{\isaliteral{3B}{\isacharsemicolon}}\isanewline
\ \ \ \ \isafun{set{\isaliteral{5F}{\isacharunderscore}}queues}\ {\isaliteral{28}{\isacharparenleft}}qs{\isaliteral{28}{\isacharparenleft}}p\ {\isaliteral{3A}{\isacharcolon}}{\isaliteral{3D}{\isacharequal}}\ i{\isaliteral{5C3C63646F743E}{\isasymcdot}}q{\isaliteral{29}{\isacharparenright}}{\isaliteral{29}{\isacharparenright}}\isanewline
\isafun{od}%
\end{isabelle}
\end{minipage}
\caption{A specification for an example tcb allocation function.\label{fig:new-tcb-def}}
\end{figure}

To illustrate function annotations, we introduce a monadic specification
for a simple function that might be used in an operating system. An imperative
program may be specified as a nondeterministic state monad~\cite{Cock_KS_08}
by defining a record containing a field for each global
variable in the program, in addition to relevant pieces of
global state. The program is then formalized as a function
that takes a state record as an input and yields an updated
record, representing global modifications, and a return value.
Nondeterministic computations yield a set of return values and state
pairs, indicating all possible ways the function could resolve
its nondeterminism.
We use ``do-notation" to phrase these monadic specifications in 
an imperative style.

Shown in \autoref{fig:new-tcb-def} \isa{\isafun{new{\isaliteral{5F}{\isacharunderscore}}tcb}} creates and enqueues
a new tcb (thread control block) kernel object. First,
in \isa{\isafun{alloc}}, an identifier is allocated
by selecting one out of the free pool
\isa{\isafun{ids}}, removing it from the pool and then returning.
Next, \isa{\isafun{create{\isaliteral{5F}{\isacharunderscore}}tcb}} creates a tcb with the appropriate
priority. \isa{\isafun{init{\isaliteral{5F}{\isacharunderscore}}tcb}} then associates this new tcb
with the previously allocated identifier in the \isa{\isafun{tcbs}}
partial map. Finally \isa{\isafun{enqueue{\isaliteral{5F}{\isacharunderscore}}tcb}} adds the identifier
to the head of the appropriate priority queue in \isa{\isafun{queues}}.
Hence there are 3 pieces of state that are relevant to the
behaviour of \isa{\isafun{new{\isaliteral{5F}{\isacharunderscore}}tcb}}: \isa{\isafun{ids}}, \isa{\isafun{tcbs}}
and \isa{\isafun{queues}}. These are represented as fields in a
state record, where \isa{\isafun{gets}\ \isafun{x}} returns the \isa{\isafun{x}}
field of the state and \isa{\isafun{set{\isaliteral{5F}{\isacharunderscore}}x}} sets that field. \isa{\isafun{return}}
simply returns the given expression, leaving the state unmodified. 
Given a set \isa{S}, \isa{\isafun{select}\ S} nondeterministically selects
a value from \isa{S}.

\begin{figure}[tb]
\begin{minipage}[t]{1\textwidth}\small
\begin{tabular}{l@ {~~\isa{{\isaliteral{5C3C65717569763E}{\isasymequiv}}}~~}l}
\isa{\isafun{valid{\isaliteral{5F}{\isacharunderscore}}id}\ i\ s} & \isa{i\ {\isaliteral{5C3C696E3E}{\isasymin}}\ \isafun{ids}\ s\ {\isaliteral{5C3C6C6F6E676C65667472696768746172726F773E}{\isasymlongleftrightarrow}}\ i\ {\isaliteral{5C3C6E6F74696E3E}{\isasymnotin}}\ \isafun{dom}\ {\isaliteral{28}{\isacharparenleft}}\isafun{tcbs}\ s{\isaliteral{29}{\isacharparenright}}}\\
\isa{\isafun{valid{\isaliteral{5F}{\isacharunderscore}}free}\ s} & \isa{{\isaliteral{5C3C666F72616C6C3E}{\isasymforall}}id{\isaliteral{2E}{\isachardot}}\ \isafun{valid{\isaliteral{5F}{\isacharunderscore}}id}\ id\ s}\\
\isa{\isafun{valid{\isaliteral{5F}{\isacharunderscore}}free{\isaliteral{5F}{\isacharunderscore}}except}\ i\ s} & \isa{{\isaliteral{5C3C666F72616C6C3E}{\isasymforall}}id{\isaliteral{2E}{\isachardot}}\ {\isaliteral{28}{\isacharparenleft}}id\ {\isaliteral{5C3C6E6F7465713E}{\isasymnoteq}}\ i\ {\isaliteral{5C3C6C6F6E6772696768746172726F773E}{\isasymlongrightarrow}}\ \isafun{valid{\isaliteral{5F}{\isacharunderscore}}id}\ id\ s{\isaliteral{29}{\isacharparenright}}\ {\isaliteral{5C3C616E643E}{\isasymand}}\ i\ {\isaliteral{5C3C6E6F74696E3E}{\isasymnotin}}\ \isafun{dom}\ {\isaliteral{28}{\isacharparenleft}}\isafun{tcbs}\ s{\isaliteral{29}{\isacharparenright}}\ {\isaliteral{5C3C616E643E}{\isasymand}}\ i\ {\isaliteral{5C3C6E6F74696E3E}{\isasymnotin}}\ \isafun{ids}\ s}
\end{tabular}
\end{minipage}
\caption{Describing a valid set of free identifiers.\label{fig:valid-free-def}}
\end{figure}

To verify this function, we may wish to first reason that \isa{\isafun{new{\isaliteral{5F}{\isacharunderscore}}tcb}} preserves the validity of the pool of free identifiers
\isa{\isafun{ids}} with respect to \isa{\isafun{tcbs}}. We introduce
an invariant \isa{\isafun{valid{\isaliteral{5F}{\isacharunderscore}}free}}, defined in \autoref{fig:valid-free-def}, which states that an identifier
is free iff it is not in the domain of \isa{\isafun{tcbs}}. In other words,
no element of \isa{\isafun{ids}} points to a tcb and all free identifiers
are necessarily in \isa{\isafun{ids}}.

We describe the preservation of this invariant as the following Hoare triple:
\begin{equation}
\isa{{\isaliteral{5C3C6C62726163653E}{\isasymlbrace}}\isafun{valid{\isaliteral{5F}{\isacharunderscore}}free}{\isaliteral{5C3C7262726163653E}{\isasymrbrace}}\ \isafun{new{\isaliteral{5F}{\isacharunderscore}}tcb}\ p\ {\isaliteral{5C3C6C62726163653E}{\isasymlbrace}}{\isaliteral{5C3C6C616D6264613E}{\isasymlambda}}{\isaliteral{5F}{\isacharunderscore}}{\isaliteral{2E}{\isachardot}}\ \isafun{valid{\isaliteral{5F}{\isacharunderscore}}free}{\isaliteral{5C3C7262726163653E}{\isasymrbrace}}} \label{eqn:new-tcb-valid-free}
\end{equation}
It states that if, before \isa{\isafun{new{\isaliteral{5F}{\isacharunderscore}}tcb}} runs, the pool
of identifiers is valid then it remains valid afterwards.
Here \isa{{\isaliteral{5C3C6C616D6264613E}{\isasymlambda}}{\isaliteral{5F}{\isacharunderscore}}{\isaliteral{2E}{\isachardot}}} binds the return value in the 
post condition to a dummy variable,
effectively ignoring it.
To prove this triple, it is sufficient to show that the operations
of \isa{\isafun{new{\isaliteral{5F}{\isacharunderscore}}tcb}} satisfy a collection of Hoare triples 
that can be composed together.
\begin{equation}
\begin{tabular}{r@ {\hspace{2 pt}}l@ {\hspace{2 pt}}r}
\isa{{\isaliteral{5C3C6C62726163653E}{\isasymlbrace}}\isafun{valid{\isaliteral{5F}{\isacharunderscore}}free}{\isaliteral{5C3C7262726163653E}{\isasymrbrace}}}
&\isa{\isafun{alloc}}
\isa{{\isaliteral{5C3C6C62726163653E}{\isasymlbrace}}{\isaliteral{5C3C6C616D6264613E}{\isasymlambda}}i\ s{\isaliteral{2E}{\isachardot}}\ \isafun{valid{\isaliteral{5F}{\isacharunderscore}}free{\isaliteral{5F}{\isacharunderscore}}except}\ i\ s{\isaliteral{5C3C7262726163653E}{\isasymrbrace}}}&\\
\isa{{\isaliteral{5C3C6C62726163653E}{\isasymlbrace}}\isafun{valid{\isaliteral{5F}{\isacharunderscore}}free{\isaliteral{5F}{\isacharunderscore}}except}\ i{\isaliteral{5C3C7262726163653E}{\isasymrbrace}}}
&\isa{\isafun{create{\isaliteral{5F}{\isacharunderscore}}tcb}\ p}
\isa{{\isaliteral{5C3C6C62726163653E}{\isasymlbrace}}{\isaliteral{5C3C6C616D6264613E}{\isasymlambda}}{\isaliteral{5F}{\isacharunderscore}}{\isaliteral{2E}{\isachardot}}\ \isafun{valid{\isaliteral{5F}{\isacharunderscore}}free{\isaliteral{5F}{\isacharunderscore}}except}\ i{\isaliteral{5C3C7262726163653E}{\isasymrbrace}}}&\\
\isa{{\isaliteral{5C3C6C62726163653E}{\isasymlbrace}}\isafun{valid{\isaliteral{5F}{\isacharunderscore}}free{\isaliteral{5F}{\isacharunderscore}}except}\ i{\isaliteral{5C3C7262726163653E}{\isasymrbrace}}}
&\isa{\isafun{init{\isaliteral{5F}{\isacharunderscore}}tcb}\ obj\ i}
\isa{{\isaliteral{5C3C6C62726163653E}{\isasymlbrace}}{\isaliteral{5C3C6C616D6264613E}{\isasymlambda}}{\isaliteral{5F}{\isacharunderscore}}{\isaliteral{2E}{\isachardot}}\ \isafun{valid{\isaliteral{5F}{\isacharunderscore}}free}{\isaliteral{5C3C7262726163653E}{\isasymrbrace}}}&\\
\isa{{\isaliteral{5C3C6C62726163653E}{\isasymlbrace}}\isafun{valid{\isaliteral{5F}{\isacharunderscore}}free}{\isaliteral{5C3C7262726163653E}{\isasymrbrace}}}
&\isa{\isafun{enqueue{\isaliteral{5F}{\isacharunderscore}}tcb}\ i\ p}
\isa{{\isaliteral{5C3C6C62726163653E}{\isasymlbrace}}{\isaliteral{5C3C6C616D6264613E}{\isasymlambda}}{\isaliteral{5F}{\isacharunderscore}}{\isaliteral{2E}{\isachardot}}\ \isafun{valid{\isaliteral{5F}{\isacharunderscore}}free}{\isaliteral{5C3C7262726163653E}{\isasymrbrace}}}&
\end{tabular} \label{eqn:hoare-collection}
\end{equation}
This demonstrates a temporary violation of \isa{\isafun{valid{\isaliteral{5F}{\isacharunderscore}}free}}
between the allocation and initialization of an identifier \isa{i}, which
is characterized by \isa{\isafun{valid{\isaliteral{5F}{\isacharunderscore}}free{\isaliteral{5F}{\isacharunderscore}}except}\ i}. 

In this small example it's clear that re-establishing \isa{\isafun{valid{\isaliteral{5F}{\isacharunderscore}}free{\isaliteral{5F}{\isacharunderscore}}except}\ i} just prior to \isa{\isafun{init{\isaliteral{5F}{\isacharunderscore}}tcb}}
in a future proof would be a trivial application of existing Hoare triples.
In the context of real software verification, where preconditions
become large collections of properties, re-establishing
the correctness of these intermediate states can result in large pieces of duplicated proof.
Moreover, for large functions it is not practical to write an individual lemma establishing
these predicates for every point in the function. Additionally
there is no established mechanism in Hoare logic for phrasing
such a lemma without manually adding assertions to the program text. In the next section we will demonstrate how
we can generate a function annotation from this proof which establishes
these intermediate properties, making them available for re-use during later
proofs over the same function.

\subsection*{Monadic Hoare Logic}
Formally,
a Hoare triple over a nondeterministic state monad is defined as follows:
\begin{equation}
\isa{{\isaliteral{5C3C6C62726163653E}{\isasymlbrace}}P{\isaliteral{5C3C7262726163653E}{\isasymrbrace}}\ f\ {\isaliteral{5C3C6C62726163653E}{\isasymlbrace}}Q{\isaliteral{5C3C7262726163653E}{\isasymrbrace}}\ {\isaliteral{5C3C65717569763E}{\isasymequiv}}\ {\isaliteral{5C3C666F72616C6C3E}{\isasymforall}}s{\isaliteral{2E}{\isachardot}}\ P\ s\ {\isaliteral{5C3C6C6F6E6772696768746172726F773E}{\isasymlongrightarrow}}\ {\isaliteral{28}{\isacharparenleft}}{\isaliteral{5C3C666F72616C6C3E}{\isasymforall}}{\isaliteral{28}{\isacharparenleft}}r{\isaliteral{2C}{\isacharcomma}}\ s{\isaliteral{27}{\isacharprime}}{\isaliteral{29}{\isacharparenright}}{\isaliteral{5C3C696E3E}{\isasymin}}f\ s{\isaliteral{2E}{\isachardot}}\ Q\ r\ s{\isaliteral{27}{\isacharprime}}{\isaliteral{29}{\isacharparenright}}} \label{eqn:valid-def}
\end{equation}
This states that a function \isa{f} satisfies a Hoare triple
if, from all states \isa{s} satisfying \isa{P\ s}, we have that
all possible computation paths of \isa{f} satisfy \isa{Q\ r\ s{\isaliteral{27}{\isacharprime}}},
where \isa{r} is the return value of \isa{f}.

To combine the results in~\eqref{eqn:hoare-collection} we use the \emph{split rule}, which states
that the precondition of a function can be used as the postcondition
of the previous function.
\begin{center}
\isa{\mbox{}\inferrule{\mbox{{\isaliteral{5C3C666F72616C6C3E}{\isasymforall}}x{\isaliteral{2E}{\isachardot}}\ {\isaliteral{5C3C6C62726163653E}{\isasymlbrace}}B\ x{\isaliteral{5C3C7262726163653E}{\isasymrbrace}}\ g\ x\ {\isaliteral{5C3C6C62726163653E}{\isasymlbrace}}C{\isaliteral{5C3C7262726163653E}{\isasymrbrace}}}\\\ \mbox{{\isaliteral{5C3C6C62726163653E}{\isasymlbrace}}A{\isaliteral{5C3C7262726163653E}{\isasymrbrace}}\ f\ {\isaliteral{5C3C6C62726163653E}{\isasymlbrace}}B{\isaliteral{5C3C7262726163653E}{\isasymrbrace}}}}{\mbox{{\isaliteral{5C3C6C62726163653E}{\isasymlbrace}}A{\isaliteral{5C3C7262726163653E}{\isasymrbrace}}\ \isafun{do}\ \ \ \ \ x\ {\isaliteral{5C3C6C6566746172726F773E}{\isasymleftarrow}}\ f{\isaliteral{3B}{\isacharsemicolon}}\ \ \ \ \ g\ x\ \isafun{od}\ {\isaliteral{5C3C6C62726163653E}{\isasymlbrace}}C{\isaliteral{5C3C7262726163653E}{\isasymrbrace}}}}} {\sc wp-split}
\end{center}
Where \isa{g} is quantified over possible return values of
\isa{f}.

\section{Using Annotations}
\label{s:annotations}

\begin{figure}[tb]
\begin{minipage}[t]{0.5\textwidth}\small
\begin{isabelle}%
\isafun{new{\isaliteral{5F}{\isacharunderscore}}tcb{\isaliteral{5F}{\isacharunderscore}}valid{\isaliteral{5F}{\isacharunderscore}}free{\isaliteral{5F}{\isacharunderscore}}ann}\ p\ {\isaliteral{5C3C65717569763E}{\isasymequiv}}\ \isafun{doA}\isanewline
\ \ \ \ i\ {\isaliteral{5C3C6C6566746172726F773E}{\isasymleftarrow}}\ {\isaliteral{5C3C6C62726163653E}{\isasymlbrace}}\isafun{valid{\isaliteral{5F}{\isacharunderscore}}free}{\isaliteral{5C3C7262726163653E}{\isasymrbrace}}\ \isafun{alloc}{\isaliteral{3B}{\isacharsemicolon}}\isanewline
\isaindent{\ \ \ \ }tcb\ {\isaliteral{5C3C6C6566746172726F773E}{\isasymleftarrow}}\ {\isaliteral{5C3C6C62726163653E}{\isasymlbrace}}\isafun{valid{\isaliteral{5F}{\isacharunderscore}}free{\isaliteral{5F}{\isacharunderscore}}except}\ i{\isaliteral{5C3C7262726163653E}{\isasymrbrace}}\ \isafun{create{\isaliteral{5F}{\isacharunderscore}}tcb}\ p{\isaliteral{3B}{\isacharsemicolon}}\isanewline
\isaindent{\ \ \ \ }{\isaliteral{5C3C6C62726163653E}{\isasymlbrace}}\isafun{valid{\isaliteral{5F}{\isacharunderscore}}free{\isaliteral{5F}{\isacharunderscore}}except}\ i{\isaliteral{5C3C7262726163653E}{\isasymrbrace}}\ \isafun{init{\isaliteral{5F}{\isacharunderscore}}tcb}\ tcb\ i{\isaliteral{3B}{\isacharsemicolon}}\isanewline
\isaindent{\ \ \ \ }{\isaliteral{5C3C6C62726163653E}{\isasymlbrace}}\isafun{valid{\isaliteral{5F}{\isacharunderscore}}free}{\isaliteral{5C3C7262726163653E}{\isasymrbrace}}\ \isafun{enqueue{\isaliteral{5F}{\isacharunderscore}}tcb}\ i\ p{\isaliteral{3B}{\isacharsemicolon}}\isanewline
\ \ \ \ {\isaliteral{5C3C6C62726163653E}{\isasymlbrace}}\isafun{valid{\isaliteral{5F}{\isacharunderscore}}free}{\isaliteral{5C3C7262726163653E}{\isasymrbrace}}\ \isafun{return}\ i\isanewline
\isafun{odA}%
\end{isabelle}
\end{minipage}
\caption{An annotation of \isa{\isafun{new{\isaliteral{5F}{\isacharunderscore}}tcb}} created during the proof of \isa{\isafun{valid{\isaliteral{5F}{\isacharunderscore}}free}}.\label{fig:new-tcb-valid-free-annotated}}
\end{figure}

The function annotation framework is designed to be
effectively transparent with respect to Hoare logic,
which enables the re-use of these intermediate properties while
requiring minimal changes to existing proofs. The goal annotation
\isa{\isafun{new{\isaliteral{5F}{\isacharunderscore}}tcb{\isaliteral{5F}{\isacharunderscore}}valid{\isaliteral{5F}{\isacharunderscore}}free{\isaliteral{5F}{\isacharunderscore}}ann}} is shown in \autoref{fig:new-tcb-valid-free-annotated}. The first step is to 
re-phrase the top level Hoare triple as an \emph{annotator}:
\begin{equation}
\isa{{\isaliteral{5C3C6C62726163653E}{\isasymlbrace}}\isafun{valid{\isaliteral{5F}{\isacharunderscore}}free}{\isaliteral{5C3C7262726163653E}{\isasymrbrace}}\ \isafun{new{\isaliteral{5F}{\isacharunderscore}}tcb}\ p\ {\isaliteral{5C3C6C62726163653E}{\isasymlbrace}}{\isaliteral{5C3C6C616D6264613E}{\isasymlambda}}{\isaliteral{5F}{\isacharunderscore}}{\isaliteral{2E}{\isachardot}}\ \isafun{valid{\isaliteral{5F}{\isacharunderscore}}free}{\isaliteral{5C3C7262726163653E}{\isasymrbrace}}\ {\isaliteral{5C3C6C616E676C653E}{\isasymlangle}}\isafun{new{\isaliteral{5F}{\isacharunderscore}}tcb{\isaliteral{5F}{\isacharunderscore}}valid{\isaliteral{5F}{\isacharunderscore}}free{\isaliteral{5F}{\isacharunderscore}}ann}\ p{\isaliteral{5C3C72616E676C653E}{\isasymrangle}}} \label{eqn:new-tcb-valid-free-2}
\end{equation}
Which states that, in addition to satisfying~\eqref{eqn:new-tcb-valid-free},
\isa{\isafun{new{\isaliteral{5F}{\isacharunderscore}}tcb}} adheres to the annotations given in \isa{\isafun{new{\isaliteral{5F}{\isacharunderscore}}tcb{\isaliteral{5F}{\isacharunderscore}}valid{\isaliteral{5F}{\isacharunderscore}}free{\isaliteral{5F}{\isacharunderscore}}ann}} given \isa{\isafun{valid{\isaliteral{5F}{\isacharunderscore}}free}} as a precondition. The additional proof obligations, showing that these annotations are satisfied, are trivially shown as a result of the existing proof.
An existing proof of~\eqref{eqn:new-tcb-valid-free} can
therefore be modified to instead show this result by mechanically exchanging
Hoare logic rules for analogous annotator rules.
In \autoref{s:annotations2} we will see how these annotations can be automatically created by Isabelle, rather than having to be explicitly specified.

The annotation itself is a monad which tracks an additional piece of
state: a boolean which indicates annotation failure. The annotation
is checked at every step of the computation, but does not affect
the behaviour of the underlying function. This annotation can therefore
be reasoned against as if it were the function it annotates. Additionally,
assuming some precondition, if we can show that no annotation will fail
then we may assume the properties stated in the annotation. 

To reason
about an annotation we introduce an \emph{annotated triple}:
\begin{center}
\isa{{\isaliteral{5C3C706172616C6C656C3E}{\isasymparallel}}P{\isaliteral{5C3C706172616C6C656C3E}{\isasymparallel}}\ F\ {\isaliteral{5C3C706172616C6C656C3E}{\isasymparallel}}Q{\isaliteral{5C3C706172616C6C656C3E}{\isasymparallel}}\ {\isaliteral{5C3C65717569763E}{\isasymequiv}}\ {\isaliteral{5C3C666F72616C6C3E}{\isasymforall}}s{\isaliteral{2E}{\isachardot}}\ {\isaliteral{5C3C6E6F743E}{\isasymnot}}\ \isafun{afails}\ {\isaliteral{28}{\isacharparenleft}}F\ s{\isaliteral{29}{\isacharparenright}}\ {\isaliteral{5C3C6C6F6E6772696768746172726F773E}{\isasymlongrightarrow}}\ P\ s\ {\isaliteral{5C3C6C6F6E6772696768746172726F773E}{\isasymlongrightarrow}}\ {\isaliteral{28}{\isacharparenleft}}{\isaliteral{5C3C666F72616C6C3E}{\isasymforall}}{\isaliteral{28}{\isacharparenleft}}r{\isaliteral{2C}{\isacharcomma}}\ s{\isaliteral{27}{\isacharprime}}{\isaliteral{29}{\isacharparenright}}{\isaliteral{5C3C696E3E}{\isasymin}}\isafun{dropA}\ F\ s{\isaliteral{2E}{\isachardot}}\ Q\ r\ s{\isaliteral{27}{\isacharprime}}{\isaliteral{29}{\isacharparenright}}} \label{eqn:validA-def}
\end{center}
Here, \isa{F} is an annotation like \isa{\isafun{new{\isaliteral{5F}{\isacharunderscore}}tcb{\isaliteral{5F}{\isacharunderscore}}valid{\isaliteral{5F}{\isacharunderscore}}free{\isaliteral{5F}{\isacharunderscore}}ann}}
and \isa{P} and \isa{Q} are pre and post conditions respectively.
\isa{\isafun{dropA}\ F} is the underlying function that \isa{F}
annotates and \isa{\isafun{afails}\ {\isaliteral{28}{\isacharparenleft}}F\ s{\isaliteral{29}{\isacharparenright}}} is true whenever \isa{F}
has an annotation that is not satisfied when proceeding from \isa{s}.
This is similar to a standard Hoare triple, with the exception that it
may take non-failure of the annotation for granted. Once an 
annotation is shown to be satisfied (using an annotator as in~\eqref{eqn:new-tcb-valid-free-2})
we can carry a result from an annotated triple over it to the underlying function.
Proving an annotated triple over an annotation is necessarily more straightforward
than a standard Hoare triple, because all the individual assertions about the intermediate
states may now be assumed to hold.

\begin{figure}[tb]
\begin{minipage}[t]{1\textwidth}\small
\begin{tabular}{l@ {~~\isa{{\isaliteral{5C3C65717569763E}{\isasymequiv}}}~~}l}
\isa{\isafun{valid{\isaliteral{5F}{\isacharunderscore}}queues}\ s} & \isa{{\isaliteral{5C3C666F72616C6C3E}{\isasymforall}}p{\isaliteral{2E}{\isachardot}}\ {\isaliteral{5C3C666F72616C6C3E}{\isasymforall}}i{\isaliteral{5C3C696E3E}{\isasymin}}\isafun{queues}\ s\ p{\isaliteral{2E}{\isachardot}}\ \isafun{tcb{\isaliteral{5F}{\isacharunderscore}}at{\isaliteral{5F}{\isacharunderscore}}prio}\ i\ p\ s}\\
\isa{\isafun{tcb{\isaliteral{5F}{\isacharunderscore}}at{\isaliteral{5F}{\isacharunderscore}}prio}\ i\ p\ s} & \isa{i\ {\isaliteral{5C3C696E3E}{\isasymin}}\ \isafun{dom}\ {\isaliteral{28}{\isacharparenleft}}\isafun{tcbs}\ s{\isaliteral{29}{\isacharparenright}}\ {\isaliteral{5C3C616E643E}{\isasymand}}\ \isafun{priority}\ {\isaliteral{28}{\isacharparenleft}}\isafun{the}\ {\isaliteral{28}{\isacharparenleft}}\isafun{tcbs}\ s\ i{\isaliteral{29}{\isacharparenright}}{\isaliteral{29}{\isacharparenright}}\ {\isaliteral{3D}{\isacharequal}}\ p}\\
\end{tabular}
\end{minipage}
\caption{Describing the correctness of scheduler queues.\label{fig:valid-queues-def}}
\end{figure}

To see how annotations are used, we introduce another
invariant \isa{\isafun{valid{\isaliteral{5F}{\isacharunderscore}}queues}} shown in \autoref{fig:valid-queues-def}. It states
that, for every scheduler queue, all the identifiers in that queue
correspond to a tcb with the appropriate priority. 
We can see that we will require \isa{\isafun{valid{\isaliteral{5F}{\isacharunderscore}}free}} as a precondition
if \isa{\isafun{new{\isaliteral{5F}{\isacharunderscore}}tcb}} is to preserve \isa{\isafun{valid{\isaliteral{5F}{\isacharunderscore}}queues}},
otherwise \isa{\isafun{alloc}} could select an identifier
that is already enqueued. We can therefore make use
of the annotation \isa{\isafun{new{\isaliteral{5F}{\isacharunderscore}}tcb{\isaliteral{5F}{\isacharunderscore}}valid{\isaliteral{5F}{\isacharunderscore}}free{\isaliteral{5F}{\isacharunderscore}}ann}} shown previously.
The goal Hoare triple is as follows:
\begin{center}
\isa{{\isaliteral{5C3C6C62726163653E}{\isasymlbrace}}\isafun{valid{\isaliteral{5F}{\isacharunderscore}}queues}\ and\ \isafun{valid{\isaliteral{5F}{\isacharunderscore}}free}{\isaliteral{5C3C7262726163653E}{\isasymrbrace}}\ \isafun{new{\isaliteral{5F}{\isacharunderscore}}tcb}\ p\ {\isaliteral{5C3C6C62726163653E}{\isasymlbrace}}{\isaliteral{5C3C6C616D6264613E}{\isasymlambda}}{\isaliteral{5F}{\isacharunderscore}}{\isaliteral{2E}{\isachardot}}\ \isafun{valid{\isaliteral{5F}{\isacharunderscore}}queues}{\isaliteral{5C3C7262726163653E}{\isasymrbrace}}} \label{eqn:new-tcb-valid-queues}
\end{center}
However, as a result of~\eqref{eqn:new-tcb-valid-free-2} which shows
that \isa{\isafun{new{\isaliteral{5F}{\isacharunderscore}}tcb}} satisfies annotations related to 
\isa{\isafun{valid{\isaliteral{5F}{\isacharunderscore}}free}}, we can instead prove this annotated triple: 
\footnote{This is formally justified by the rule~\eqref{eqn:use-annotated-atomized} introduced later in \autoref{s:annotations2}} 
\begin{center}
\isa{{\isaliteral{5C3C706172616C6C656C3E}{\isasymparallel}}\isafun{valid{\isaliteral{5F}{\isacharunderscore}}queues}\ and\ \isafun{valid{\isaliteral{5F}{\isacharunderscore}}free}{\isaliteral{5C3C706172616C6C656C3E}{\isasymparallel}}\ \isafun{new{\isaliteral{5F}{\isacharunderscore}}tcb{\isaliteral{5F}{\isacharunderscore}}valid{\isaliteral{5F}{\isacharunderscore}}free{\isaliteral{5F}{\isacharunderscore}}ann}\ p\ {\isaliteral{5C3C706172616C6C656C3E}{\isasymparallel}}{\isaliteral{5C3C6C616D6264613E}{\isasymlambda}}{\isaliteral{5F}{\isacharunderscore}}{\isaliteral{2E}{\isachardot}}\ \isafun{valid{\isaliteral{5F}{\isacharunderscore}}queues}{\isaliteral{5C3C706172616C6C656C3E}{\isasymparallel}}} \label{eqn:new-tcb-valid-queues-annotated}
\end{center}
We prove these sorts of annotated triples by decomposing
them with an analogous rule to {\sc wp-split} and then converting them into ordinary Hoare triples.
The annotation on a function \isa{{\isaliteral{5C3C6C62726163653E}{\isasymlbrace}}P{\isaliteral{5C3C7262726163653E}{\isasymrbrace}}\ f} can be read as "f, given P".
The following rule allows such an annotation in an annotated triple to be
assumed as a precondition, while converting into a standard Hoare 
triple:
\begin{equation}
\isa{\mbox{}\inferrule{\mbox{{\isaliteral{5C3C6C62726163653E}{\isasymlbrace}}R\ and\ P{\isaliteral{5C3C7262726163653E}{\isasymrbrace}}\ f\ {\isaliteral{5C3C6C62726163653E}{\isasymlbrace}}Q{\isaliteral{5C3C7262726163653E}{\isasymrbrace}}}}{\mbox{{\isaliteral{5C3C706172616C6C656C3E}{\isasymparallel}}R{\isaliteral{5C3C706172616C6C656C3E}{\isasymparallel}}\ {\isaliteral{5C3C6C62726163653E}{\isasymlbrace}}P{\isaliteral{5C3C7262726163653E}{\isasymrbrace}}\ f\ {\isaliteral{5C3C706172616C6C656C3E}{\isasymparallel}}Q{\isaliteral{5C3C706172616C6C656C3E}{\isasymparallel}}}}} \label{eqn:valid-annotated-atomized}
\end{equation}
This allows a standard Hoare triple to be applied as if that
precondition was established in a traditional Hoare logic proof.

A strong splitting rule, which combines the annotated triple splitting rule 
and~\eqref{eqn:valid-annotated-atomized}, can be directly applied to decompose
an annotated triple into a collection of standard Hoare triples that are strengthened with
annotations.
\begin{center}
\isa{\mbox{}\inferrule{\mbox{{\isaliteral{5C3C666F72616C6C3E}{\isasymforall}}x{\isaliteral{2E}{\isachardot}}\ {\isaliteral{5C3C706172616C6C656C3E}{\isasymparallel}}B\ x{\isaliteral{5C3C706172616C6C656C3E}{\isasymparallel}}\ G\ x\ {\isaliteral{5C3C706172616C6C656C3E}{\isasymparallel}}C{\isaliteral{5C3C706172616C6C656C3E}{\isasymparallel}}}\\\ \mbox{{\isaliteral{5C3C6C62726163653E}{\isasymlbrace}}A\ and\ P{\isaliteral{5C3C7262726163653E}{\isasymrbrace}}\ f\ {\isaliteral{5C3C6C62726163653E}{\isasymlbrace}}B{\isaliteral{5C3C7262726163653E}{\isasymrbrace}}}}{\mbox{{\isaliteral{5C3C706172616C6C656C3E}{\isasymparallel}}A{\isaliteral{5C3C706172616C6C656C3E}{\isasymparallel}}\ \isafun{doA}\ \ \ \ \ x\ {\isaliteral{5C3C6C6566746172726F773E}{\isasymleftarrow}}\ {\isaliteral{5C3C6C62726163653E}{\isasymlbrace}}P{\isaliteral{5C3C7262726163653E}{\isasymrbrace}}\ f{\isaliteral{3B}{\isacharsemicolon}}\ \ \ \ \ G\ x\ \isafun{odA}\ {\isaliteral{5C3C706172616C6C656C3E}{\isasymparallel}}C{\isaliteral{5C3C706172616C6C656C3E}{\isasymparallel}}}}} {\sc wp-strong-split}
\end{center}
To show that \isa{\isafun{new{\isaliteral{5F}{\isacharunderscore}}tcb}} preserves \isa{\isafun{valid{\isaliteral{5F}{\isacharunderscore}}queues}} we first
establish the precondition for \isa{\isafun{enqueue{\isaliteral{5F}{\isacharunderscore}}tcb}} to preserve \isa{\isafun{valid{\isaliteral{5F}{\isacharunderscore}}queues}}:
\begin{center}
\isa{{\isaliteral{5C3C6C62726163653E}{\isasymlbrace}}\isafun{valid{\isaliteral{5F}{\isacharunderscore}}queues}\ and\ \isafun{tcb{\isaliteral{5F}{\isacharunderscore}}at{\isaliteral{5F}{\isacharunderscore}}prio}\ i\ p{\isaliteral{5C3C7262726163653E}{\isasymrbrace}}\ \isafun{enqueue{\isaliteral{5F}{\isacharunderscore}}tcb}\ i\ p\ {\isaliteral{5C3C6C62726163653E}{\isasymlbrace}}{\isaliteral{5C3C6C616D6264613E}{\isasymlambda}}{\isaliteral{5F}{\isacharunderscore}}{\isaliteral{2E}{\isachardot}}\ \isafun{valid{\isaliteral{5F}{\isacharunderscore}}queues}{\isaliteral{5C3C7262726163653E}{\isasymrbrace}}} \label{eqn:enqueue-tcb-valid-queues}
\end{center}
We can then demonstrate that \isa{\isafun{init{\isaliteral{5F}{\isacharunderscore}}tcb}} preserves \isa{\isafun{valid{\isaliteral{5F}{\isacharunderscore}}queues}}
assuming the identifier being initialized is not already enqueued:
\begin{center}
\isa{{\isaliteral{5C3C6C62726163653E}{\isasymlbrace}}\isafun{valid{\isaliteral{5F}{\isacharunderscore}}queues}\ and\ \isafun{not{\isaliteral{5F}{\isacharunderscore}}queued}\ i{\isaliteral{5C3C7262726163653E}{\isasymrbrace}}\ \isafun{init{\isaliteral{5F}{\isacharunderscore}}tcb}\ obj\ i\ {\isaliteral{5C3C6C62726163653E}{\isasymlbrace}}{\isaliteral{5C3C6C616D6264613E}{\isasymlambda}}{\isaliteral{5F}{\isacharunderscore}}{\isaliteral{2E}{\isachardot}}\ \isafun{valid{\isaliteral{5F}{\isacharunderscore}}queues}{\isaliteral{5C3C7262726163653E}{\isasymrbrace}}} \label{eqn:init-tcb-valid-queues}
\end{center}
where \isa{\isafun{not{\isaliteral{5F}{\isacharunderscore}}queued}\ i\ s\ {\isaliteral{5C3C65717569763E}{\isasymequiv}}\ {\isaliteral{5C3C666F72616C6C3E}{\isasymforall}}p{\isaliteral{2E}{\isachardot}}\ i\ {\isaliteral{5C3C6E6F74696E3E}{\isasymnotin}}\ \isafun{queues}\ s\ p}. If applied directly, this will require additional reasoning that \isa{\isafun{create{\isaliteral{5F}{\isacharunderscore}}tcb}}
preserves the fact that the identifier is not enqueued and that \isa{\isafun{alloc}}
selects an identifier that is not enqueued. Recall, however, that this proof
is over an annotated function, and at this point we have that \isa{\isafun{valid{\isaliteral{5F}{\isacharunderscore}}free{\isaliteral{5F}{\isacharunderscore}}except}\ i} holds.
{\sc wp-strong-split} therefore will have added \isa{\isafun{valid{\isaliteral{5F}{\isacharunderscore}}free{\isaliteral{5F}{\isacharunderscore}}except}\ i} to the precondition
for \isa{\isafun{create{\isaliteral{5F}{\isacharunderscore}}tcb}}. Using the implication \isa{\isafun{valid{\isaliteral{5F}{\isacharunderscore}}queues}\ s\ {\isaliteral{5C3C616E643E}{\isasymand}}\ \isafun{valid{\isaliteral{5F}{\isacharunderscore}}free{\isaliteral{5F}{\isacharunderscore}}except}\ i\ s\ {\isaliteral{5C3C6C6F6E6772696768746172726F773E}{\isasymlongrightarrow}}\ \isafun{not{\isaliteral{5F}{\isacharunderscore}}queued}\ i\ s}
we may strengthen the precondition to instead assume \isa{\isafun{valid{\isaliteral{5F}{\isacharunderscore}}free{\isaliteral{5F}{\isacharunderscore}}except}\ i}:
\begin{center}
\isa{{\isaliteral{5C3C6C62726163653E}{\isasymlbrace}}\isafun{valid{\isaliteral{5F}{\isacharunderscore}}queues}\ and\ \isafun{valid{\isaliteral{5F}{\isacharunderscore}}free{\isaliteral{5F}{\isacharunderscore}}except}\ i{\isaliteral{5C3C7262726163653E}{\isasymrbrace}}\ \isafun{init{\isaliteral{5F}{\isacharunderscore}}tcb}\ tcb\ i\ {\isaliteral{5C3C6C62726163653E}{\isasymlbrace}}{\isaliteral{5C3C6C616D6264613E}{\isasymlambda}}{\isaliteral{5F}{\isacharunderscore}}{\isaliteral{2E}{\isachardot}}\ \isafun{valid{\isaliteral{5F}{\isacharunderscore}}queues}{\isaliteral{5C3C7262726163653E}{\isasymrbrace}}} \label{eqn:init-tcb-valid-queues-annotated}
\end{center}

This uses the annotation granted from {\sc wp-strong-split}, therefore the only precondition
that needs to be propagated is \isa{\isafun{valid{\isaliteral{5F}{\isacharunderscore}}queues}}, and
we no longer have to reason about \isa{i} not being enqueued. In a larger
function this could potentially avoid propagating this precondition up through
several operations and duplicating a significant amount of existing reasoning.
\clearpage
\begin{figure}[tb]
\begin{isabelle}%
\isafun{new{\isaliteral{5F}{\isacharunderscore}}tcb{\isaliteral{5F}{\isacharunderscore}}valid{\isaliteral{5F}{\isacharunderscore}}queues{\isaliteral{5F}{\isacharunderscore}}ann}\ p\ {\isaliteral{5C3C65717569763E}{\isasymequiv}}\ \isafun{doA}\isanewline
\ \ \ \ i\ {\isaliteral{5C3C6C6566746172726F773E}{\isasymleftarrow}}\ {\isaliteral{5C3C6C62726163653E}{\isasymlbrace}}\isafun{valid{\isaliteral{5F}{\isacharunderscore}}queues}{\isaliteral{5C3C7262726163653E}{\isasymrbrace}}\ \isafun{alloc}{\isaliteral{3B}{\isacharsemicolon}}\isanewline
\isaindent{\ \ \ \ }tcb\ {\isaliteral{5C3C6C6566746172726F773E}{\isasymleftarrow}}\ {\isaliteral{5C3C6C62726163653E}{\isasymlbrace}}\isafun{valid{\isaliteral{5F}{\isacharunderscore}}queues}{\isaliteral{5C3C7262726163653E}{\isasymrbrace}}\ \isafun{create{\isaliteral{5F}{\isacharunderscore}}tcb}\ p{\isaliteral{3B}{\isacharsemicolon}}\isanewline
\isaindent{\ \ \ \ }{\isaliteral{5C3C6C62726163653E}{\isasymlbrace}}\isafun{valid{\isaliteral{5F}{\isacharunderscore}}queues}\ and\ \isafun{not{\isaliteral{5F}{\isacharunderscore}}queued}\ i\ and\ {\isaliteral{28}{\isacharparenleft}}{\isaliteral{5C3C6C616D6264613E}{\isasymlambda}}{\isaliteral{5F}{\isacharunderscore}}{\isaliteral{2E}{\isachardot}}\ \isafun{priority}\ tcb\ {\isaliteral{3D}{\isacharequal}}\ p{\isaliteral{29}{\isacharparenright}}{\isaliteral{5C3C7262726163653E}{\isasymrbrace}}\isanewline
\isaindent{\ \ \ \ }\isafun{init{\isaliteral{5F}{\isacharunderscore}}tcb}\ tcb\ i{\isaliteral{3B}{\isacharsemicolon}}\isanewline
\isaindent{\ \ \ \ }{\isaliteral{5C3C6C62726163653E}{\isasymlbrace}}\isafun{valid{\isaliteral{5F}{\isacharunderscore}}queues}\ and\ \isafun{tcb{\isaliteral{5F}{\isacharunderscore}}at{\isaliteral{5F}{\isacharunderscore}}prio}\ i\ p{\isaliteral{5C3C7262726163653E}{\isasymrbrace}}\ \isafun{enqueue{\isaliteral{5F}{\isacharunderscore}}tcb}\ i\ p{\isaliteral{3B}{\isacharsemicolon}}\isanewline
\ \ \ \ {\isaliteral{5C3C6C62726163653E}{\isasymlbrace}}\isafun{valid{\isaliteral{5F}{\isacharunderscore}}queues}{\isaliteral{5C3C7262726163653E}{\isasymrbrace}}\ \isafun{return}\ i\isanewline
\isafun{odA}%
\end{isabelle}
\caption{Annotations for \isa{\isafun{new{\isaliteral{5F}{\isacharunderscore}}tcb}} from the proof of \isa{\isafun{valid{\isaliteral{5F}{\isacharunderscore}}queues}}.\label{fig:new-tcb-valid-queues-annotated}}
\end{figure}

\begin{figure}[tb]
\begin{isabelle}%
\isafun{new{\isaliteral{5F}{\isacharunderscore}}tcb{\isaliteral{5F}{\isacharunderscore}}valid{\isaliteral{5F}{\isacharunderscore}}free{\isaliteral{5F}{\isacharunderscore}}ann}\ p\ {\isaliteral{5C3C626F777469653E}{\isasymbowtie}}\ \isafun{new{\isaliteral{5F}{\isacharunderscore}}tcb{\isaliteral{5F}{\isacharunderscore}}valid{\isaliteral{5F}{\isacharunderscore}}queues{\isaliteral{5F}{\isacharunderscore}}ann}\ p\ {\isaliteral{3D}{\isacharequal}}\ \isafun{doA}\isanewline
\ \ \ \ i\ {\isaliteral{5C3C6C6566746172726F773E}{\isasymleftarrow}}\ {\isaliteral{5C3C6C62726163653E}{\isasymlbrace}}\isafun{valid{\isaliteral{5F}{\isacharunderscore}}free}\ and\ \isafun{valid{\isaliteral{5F}{\isacharunderscore}}queues}{\isaliteral{5C3C7262726163653E}{\isasymrbrace}}\ \isafun{alloc}{\isaliteral{3B}{\isacharsemicolon}}\isanewline
\isaindent{\ \ \ \ }tcb\ {\isaliteral{5C3C6C6566746172726F773E}{\isasymleftarrow}}\ {\isaliteral{5C3C6C62726163653E}{\isasymlbrace}}\isafun{valid{\isaliteral{5F}{\isacharunderscore}}free{\isaliteral{5F}{\isacharunderscore}}except}\ i\ and\ \isafun{valid{\isaliteral{5F}{\isacharunderscore}}queues}{\isaliteral{5C3C7262726163653E}{\isasymrbrace}}\ \isafun{create{\isaliteral{5F}{\isacharunderscore}}tcb}\ p{\isaliteral{3B}{\isacharsemicolon}}\isanewline
\isaindent{\ \ \ \ }{\isaliteral{5C3C6C62726163653E}{\isasymlbrace}}\isafun{valid{\isaliteral{5F}{\isacharunderscore}}free{\isaliteral{5F}{\isacharunderscore}}except}\ i\ and\ \isafun{valid{\isaliteral{5F}{\isacharunderscore}}queues}\ and\ \isafun{not{\isaliteral{5F}{\isacharunderscore}}queued}\ i\ and\isanewline
\isaindent{\ \ \ \ {\isaliteral{5C3C6C62726163653E}{\isasymlbrace}}}{\isaliteral{28}{\isacharparenleft}}{\isaliteral{5C3C6C616D6264613E}{\isasymlambda}}{\isaliteral{5F}{\isacharunderscore}}{\isaliteral{2E}{\isachardot}}\ \isafun{priority}\ tcb\ {\isaliteral{3D}{\isacharequal}}\ p{\isaliteral{29}{\isacharparenright}}{\isaliteral{5C3C7262726163653E}{\isasymrbrace}}\isanewline
\isaindent{\ \ \ \ }\isafun{init{\isaliteral{5F}{\isacharunderscore}}tcb}\ tcb\ i{\isaliteral{3B}{\isacharsemicolon}}\isanewline
\isaindent{\ \ \ \ }{\isaliteral{5C3C6C62726163653E}{\isasymlbrace}}\isafun{valid{\isaliteral{5F}{\isacharunderscore}}free}\ and\ \isafun{valid{\isaliteral{5F}{\isacharunderscore}}queues}\ and\ \isafun{tcb{\isaliteral{5F}{\isacharunderscore}}at{\isaliteral{5F}{\isacharunderscore}}prio}\ i\ p{\isaliteral{5C3C7262726163653E}{\isasymrbrace}}\ \isafun{enqueue{\isaliteral{5F}{\isacharunderscore}}tcb}\ i\ p{\isaliteral{3B}{\isacharsemicolon}}\isanewline
\ \ \ \ {\isaliteral{5C3C6C62726163653E}{\isasymlbrace}}\isafun{valid{\isaliteral{5F}{\isacharunderscore}}free}\ and\ \isafun{valid{\isaliteral{5F}{\isacharunderscore}}queues}{\isaliteral{5C3C7262726163653E}{\isasymrbrace}}\ \isafun{return}\ i\isanewline
\isafun{odA}%
\end{isabelle}
\caption{A combination of two annotations for \isa{\isafun{new{\isaliteral{5F}{\isacharunderscore}}tcb}}.\label{fig:new-tcb-merge}}
\end{figure}

Similar to the proof for \isa{\isafun{valid{\isaliteral{5F}{\isacharunderscore}}free}} we can create annotations
regarding \isa{\isafun{valid{\isaliteral{5F}{\isacharunderscore}}queues}} as shown in \autoref{fig:new-tcb-valid-queues-annotated}. Note that only properties specifically related to \isa{\isafun{valid{\isaliteral{5F}{\isacharunderscore}}queues}}
are used as annotations. To combine two annotated functions we define a merge operation \isa{{\isaliteral{5C3C626F777469653E}{\isasymbowtie}}}, which produces a new annotated function that is simply the conjunction of both
annotations, and can be used whenever both their preconditions are established. The result of merging both of the previous annotations for \isa{\isafun{new{\isaliteral{5F}{\isacharunderscore}}tcb}} is shown
in \autoref{fig:new-tcb-merge}.

\section{Creating Annotations}
\label{s:annotations2}

In this section we describe how function annotations were formalized
 using Isabelle and how they were incorporated into an existing Hoare logic
vcg to allow them to be seamlessly
integrated with existing proofs. Additionally we show how function
annotations can be easily generated by Isabelle and exported by using
schematic lemmas.

Function annotations are implemented by creating a new
function which, effectively, has an assertion between every operation. To
demonstrate that an annotation is valid (i.e. create an annotator 
like~\eqref{eqn:new-tcb-valid-free-2})
 it is sufficient to show that
none of these assertions will fail under a given precondition.
An extra flag is tracked, in addition to the global state record,
which indicates failure of an annotation. An annotation,
therefore, simply evaluates the function as normal but
also tests annotations.
\begin{center}
\isa{{\isaliteral{5C3C6C62726163653E}{\isasymlbrace}}P{\isaliteral{5C3C7262726163653E}{\isasymrbrace}}\ f\ {\isaliteral{3D}{\isacharequal}}\ {\isaliteral{28}{\isacharparenleft}}{\isaliteral{5C3C6C616D6264613E}{\isasymlambda}}s{\isaliteral{2E}{\isachardot}}\ {\isaliteral{28}{\isacharparenleft}}f\ s{\isaliteral{2C}{\isacharcomma}}\ {\isaliteral{5C3C6E6F743E}{\isasymnot}}\ P\ s{\isaliteral{29}{\isacharparenright}}{\isaliteral{29}{\isacharparenright}}} \label{eqn:annotate-lift-def}
\end{center}
\begin{figure}[b]
\begin{tabular}{l c l}
\begin{minipage}[t]{0.08\textwidth}
\begin{isabelle}%
\isafun{doA}\isanewline
\ \ \ \ x\ {\isaliteral{5C3C6C6566746172726F773E}{\isasymleftarrow}}\ F{\isaliteral{3B}{\isacharsemicolon}}\isanewline
\ \ \ \ G\ x\isanewline
\isafun{odA}%
\end{isabelle}
\end{minipage}
&\isa{{\isaliteral{5C3C65717569763E}{\isasymequiv}}}&
\begin{minipage}[t]{0.5\textwidth}
\begin{isabelle}%
{\isaliteral{5C3C6C616D6264613E}{\isasymlambda}}s{\isaliteral{2E}{\isachardot}}\ {\isaliteral{28}{\isacharparenleft}}{\isaliteral{28}{\isacharparenleft}}\isafun{do}\isanewline
\isaindent{{\isaliteral{5C3C6C616D6264613E}{\isasymlambda}}s{\isaliteral{2E}{\isachardot}}\ {\isaliteral{28}{\isacharparenleft}}\ }\ \ \ \ x\ {\isaliteral{5C3C6C6566746172726F773E}{\isasymleftarrow}}\ \isafun{dropA}\ F{\isaliteral{3B}{\isacharsemicolon}}\isanewline
\isaindent{{\isaliteral{5C3C6C616D6264613E}{\isasymlambda}}s{\isaliteral{2E}{\isachardot}}\ {\isaliteral{28}{\isacharparenleft}}\ }\ \ \ \ \isafun{dropA}\ {\isaliteral{28}{\isacharparenleft}}G\ x{\isaliteral{29}{\isacharparenright}}\isanewline
\isaindent{{\isaliteral{5C3C6C616D6264613E}{\isasymlambda}}s{\isaliteral{2E}{\isachardot}}\ {\isaliteral{28}{\isacharparenleft}}\ }\isafun{od}{\isaliteral{29}{\isacharparenright}}\isanewline
\isaindent{{\isaliteral{5C3C6C616D6264613E}{\isasymlambda}}s{\isaliteral{2E}{\isachardot}}\ {\isaliteral{28}{\isacharparenleft}}\ }s{\isaliteral{2C}{\isacharcomma}}\isanewline
\isaindent{{\isaliteral{5C3C6C616D6264613E}{\isasymlambda}}s{\isaliteral{2E}{\isachardot}}\ \ }\isafun{can{\isaliteral{5F}{\isacharunderscore}}fail{\isaliteral{5F}{\isacharunderscore}}from}\ F\ G\ s\ {\isaliteral{5C3C6F723E}{\isasymor}}\ \isafun{afails}\ {\isaliteral{28}{\isacharparenleft}}F\ s{\isaliteral{29}{\isacharparenright}}{\isaliteral{29}{\isacharparenright}}%
\end{isabelle}
\end{minipage} \smallskip \\
\isa{\isafun{can{\isaliteral{5F}{\isacharunderscore}}fail{\isaliteral{5F}{\isacharunderscore}}from}\ F\ G\ s}
&\isa{{\isaliteral{5C3C65717569763E}{\isasymequiv}}}&
\isa{\isafun{True}\ {\isaliteral{5C3C696E3E}{\isasymin}}\ \isafun{afails}\ {\isaliteral{60}{\isacharbackquote}}\ {\isaliteral{28}{\isacharparenleft}}{\isaliteral{5C3C6C616D6264613E}{\isasymlambda}}{\isaliteral{28}{\isacharparenleft}}x{\isaliteral{2C}{\isacharcomma}}\ y{\isaliteral{29}{\isacharparenright}}{\isaliteral{2E}{\isachardot}}\ G\ x\ y{\isaliteral{29}{\isacharparenright}}\ {\isaliteral{60}{\isacharbackquote}}\ \isafun{dropA}\ F\ s} \smallskip\\
\isa{\isafun{dropA}\ {\isaliteral{5C3C6C62726163653E}{\isasymlbrace}}P{\isaliteral{5C3C7262726163653E}{\isasymrbrace}}\ f}
&\isa{{\isaliteral{3D}{\isacharequal}}}&
\isa{f} \smallskip \\
\isa{\isafun{afails}\ {\isaliteral{28}{\isacharparenleft}}{\isaliteral{5C3C6C62726163653E}{\isasymlbrace}}P{\isaliteral{5C3C7262726163653E}{\isasymrbrace}}\ f\ s{\isaliteral{29}{\isacharparenright}}}
&\isa{{\isaliteral{3D}{\isacharequal}}}&
\isa{{\isaliteral{5C3C6E6F743E}{\isasymnot}}\ P\ s}
\end{tabular}
\caption{Composing annotated functions.\label{fig:bindA-def}}
\end{figure}
To evaluate the composition of two annotated functions we compose
their inner functions and then set the failure flag if any
possible nondeterministic branching of the composition can fail.
The definition is given in \autoref{fig:bindA-def} with
relevant lemmas to characterize \isa{\isafun{dropA}} and \isa{\isafun{afails}}.
\isa{\isafun{can{\isaliteral{5F}{\isacharunderscore}}fail{\isaliteral{5F}{\isacharunderscore}}from}\ F\ G\ s} runs \isa{G} against all possible
results of \isa{F} from \isa{s} and then tests annotation
failure on all possible outcomes.

A function is said to satisfy an annotation under some precondition
if, by asserting that precondition at the beginning of the function,
 no assertions will fail. To formalize this, we define
 a partial ordering \isa{{\isaliteral{5C3C737173756273657465713E}{\isasymsqsubseteq}}} on annotated functions that can be
described by the strength of the annotations:
\begin{center}
\isa{F\ {\isaliteral{5C3C737173756273657465713E}{\isasymsqsubseteq}}\ G\ {\isaliteral{5C3C65717569763E}{\isasymequiv}}\ {\isaliteral{5C3C666F72616C6C3E}{\isasymforall}}s{\isaliteral{2E}{\isachardot}}\ {\isaliteral{28}{\isacharparenleft}}{\isaliteral{5C3C6E6F743E}{\isasymnot}}\ \isafun{afails}\ {\isaliteral{28}{\isacharparenleft}}F\ s{\isaliteral{29}{\isacharparenright}}\ {\isaliteral{5C3C6C6F6E6772696768746172726F773E}{\isasymlongrightarrow}}\ \isafun{dropA}\ F\ s\ {\isaliteral{3D}{\isacharequal}}\ \isafun{dropA}\ G\ s{\isaliteral{29}{\isacharparenright}}\ {\isaliteral{5C3C616E643E}{\isasymand}}\ {\isaliteral{28}{\isacharparenleft}}\isafun{afails}\ {\isaliteral{28}{\isacharparenleft}}G\ s{\isaliteral{29}{\isacharparenright}}\ {\isaliteral{5C3C6C6F6E6772696768746172726F773E}{\isasymlongrightarrow}}\ \isafun{afails}\ {\isaliteral{28}{\isacharparenleft}}F\ s{\isaliteral{29}{\isacharparenright}}{\isaliteral{29}{\isacharparenright}}} \label{eqn:naf-sub-def}
\end{center}
which states that the annotation \isa{F} is stronger than \isa{G} if, under
non-failure, they annotate the same function and failure of
\isa{G} always implies failure of \isa{F}.
It is best characterized by the following rule:
\begin{center}
\isa{\mbox{}\inferrule{\mbox{{\isaliteral{5C3C666F72616C6C3E}{\isasymforall}}s{\isaliteral{2E}{\isachardot}}\ P\ s\ {\isaliteral{5C3C6C6F6E6772696768746172726F773E}{\isasymlongrightarrow}}\ Q\ s}}{\mbox{{\isaliteral{5C3C6C62726163653E}{\isasymlbrace}}P{\isaliteral{5C3C7262726163653E}{\isasymrbrace}}\ f\ {\isaliteral{5C3C737173756273657465713E}{\isasymsqsubseteq}}\ {\isaliteral{5C3C6C62726163653E}{\isasymlbrace}}Q{\isaliteral{5C3C7262726163653E}{\isasymrbrace}}\ f}}} \label{eqn:naf-sub-annotate-atomized}
\end{center}
Hence \isa{f} satisfies some annotation \isa{F} under
\isa{P} if \isa{{\isaliteral{5C3C6C62726163653E}{\isasymlbrace}}P{\isaliteral{5C3C7262726163653E}{\isasymrbrace}}\ f\ {\isaliteral{5C3C737173756273657465713E}{\isasymsqsubseteq}}\ F}.
When a function satisfies an annotation, 
one can reason about the function by reasoning about the annotation
via an annotated triple, which only evaluates
under non-failure and thus all annotations may be assumed.
\begin{equation}
\isa{\mbox{}\inferrule{\mbox{{\isaliteral{5C3C6C62726163653E}{\isasymlbrace}}P{\isaliteral{5C3C7262726163653E}{\isasymrbrace}}\ f\ {\isaliteral{5C3C737173756273657465713E}{\isasymsqsubseteq}}\ F}\\\ \mbox{{\isaliteral{5C3C706172616C6C656C3E}{\isasymparallel}}P{\isaliteral{5C3C706172616C6C656C3E}{\isasymparallel}}\ F\ {\isaliteral{5C3C706172616C6C656C3E}{\isasymparallel}}Q{\isaliteral{5C3C706172616C6C656C3E}{\isasymparallel}}}}{\mbox{{\isaliteral{5C3C6C62726163653E}{\isasymlbrace}}P{\isaliteral{5C3C7262726163653E}{\isasymrbrace}}\ f\ {\isaliteral{5C3C6C62726163653E}{\isasymlbrace}}Q{\isaliteral{5C3C7262726163653E}{\isasymrbrace}}}}} \label{eqn:use-annotated-atomized}
\end{equation}
To prove adherence to an annotation we use an annotator.
It simply states that, in addition to satisfying the given
Hoare triple, the function adheres to the given annotation. 
\begin{equation}
\isa{{\isaliteral{5C3C6C62726163653E}{\isasymlbrace}}P{\isaliteral{5C3C7262726163653E}{\isasymrbrace}}\ f\ {\isaliteral{5C3C6C62726163653E}{\isasymlbrace}}Q{\isaliteral{5C3C7262726163653E}{\isasymrbrace}}\ {\isaliteral{5C3C6C616E676C653E}{\isasymlangle}}F{\isaliteral{5C3C72616E676C653E}{\isasymrangle}}\ {\isaliteral{5C3C65717569763E}{\isasymequiv}}\ {\isaliteral{5C3C6C62726163653E}{\isasymlbrace}}P{\isaliteral{5C3C7262726163653E}{\isasymrbrace}}\ f\ {\isaliteral{5C3C6C62726163653E}{\isasymlbrace}}Q{\isaliteral{5C3C7262726163653E}{\isasymrbrace}}\ {\isaliteral{5C3C616E643E}{\isasymand}}\ {\isaliteral{5C3C6C62726163653E}{\isasymlbrace}}P{\isaliteral{5C3C7262726163653E}{\isasymrbrace}}\ f\ {\isaliteral{5C3C737173756273657465713E}{\isasymsqsubseteq}}\ F} \label{eqn:valid-annotated-def}
\end{equation}
There are two motivations for proving annotation adherence
this way. As will be explained in the next section, this
approach allows the annotation to be ``collected" by Isabelle
rather than provided explicitly by the user. Additionally
it provides a clear notion of an \emph{input} function
that is being proved against and an \emph{output} annotation
that is being produced/satisfied. In the general case
the function may itself already have annotations; by phrasing
the triple in this way we can distinguish annotations we may
use during the proof of the Hoare triple and annotations
that we are producing/satisfying.
To illustrate this, recall the proof that \isa{\isafun{new{\isaliteral{5F}{\isacharunderscore}}tcb}}
preserves \isa{\isafun{valid{\isaliteral{5F}{\isacharunderscore}}queues}}. While using
the annotation created during the proof of
\isa{\isafun{valid{\isaliteral{5F}{\isacharunderscore}}free}} we also want to produce an annotation
for \isa{\isafun{valid{\isaliteral{5F}{\isacharunderscore}}queues}}. This can be phrased with 
a strong annotator, which has an analogous definition to the one given
in~\eqref{eqn:valid-annotated-def}:
\begin{center}
\isa{{\isaliteral{5C3C706172616C6C656C3E}{\isasymparallel}}\isafun{valid{\isaliteral{5F}{\isacharunderscore}}queues}{\isaliteral{5C3C706172616C6C656C3E}{\isasymparallel}}\ \isafun{new{\isaliteral{5F}{\isacharunderscore}}tcb{\isaliteral{5F}{\isacharunderscore}}valid{\isaliteral{5F}{\isacharunderscore}}free{\isaliteral{5F}{\isacharunderscore}}ann}\ p\ {\isaliteral{5C3C706172616C6C656C3E}{\isasymparallel}}{\isaliteral{5C3C6C616D6264613E}{\isasymlambda}}{\isaliteral{5F}{\isacharunderscore}}{\isaliteral{2E}{\isachardot}}\ \isafun{valid{\isaliteral{5F}{\isacharunderscore}}queues}{\isaliteral{5C3C706172616C6C656C3E}{\isasymparallel}}\ {\isaliteral{5C3C6C616E676C653E}{\isasymlangle}}\isafun{new{\isaliteral{5F}{\isacharunderscore}}tcb{\isaliteral{5F}{\isacharunderscore}}valid{\isaliteral{5F}{\isacharunderscore}}queues{\isaliteral{5F}{\isacharunderscore}}ann}\ p{\isaliteral{5C3C72616E676C653E}{\isasymrangle}}}
\end{center}
Once this proof is complete, we now
have that \isa{\isafun{new{\isaliteral{5F}{\isacharunderscore}}tcb}}, in addition to satisfying
\isa{\isafun{new{\isaliteral{5F}{\isacharunderscore}}tcb{\isaliteral{5F}{\isacharunderscore}}valid{\isaliteral{5F}{\isacharunderscore}}free{\isaliteral{5F}{\isacharunderscore}}ann}} under \isa{\isafun{valid{\isaliteral{5F}{\isacharunderscore}}free}},
satisfies \isa{\isafun{new{\isaliteral{5F}{\isacharunderscore}}tcb{\isaliteral{5F}{\isacharunderscore}}valid{\isaliteral{5F}{\isacharunderscore}}queues{\isaliteral{5F}{\isacharunderscore}}ann}} under 
\isa{\isafun{valid{\isaliteral{5F}{\isacharunderscore}}queues}\ and\ \isafun{valid{\isaliteral{5F}{\isacharunderscore}}free}}.
To merge two annotations, as seen previously in
\autoref{fig:new-tcb-merge}, we simply take the disjunction
of their failure flags:
\begin{center}
\isa{F\ {\isaliteral{5C3C626F777469653E}{\isasymbowtie}}\ G\ {\isaliteral{5C3C65717569763E}{\isasymequiv}}\ {\isaliteral{5C3C6C616D6264613E}{\isasymlambda}}s{\isaliteral{2E}{\isachardot}}\ {\isaliteral{28}{\isacharparenleft}}\isafun{dropA}\ F\ s{\isaliteral{2C}{\isacharcomma}}\ \isafun{afails}\ {\isaliteral{28}{\isacharparenleft}}F\ s{\isaliteral{29}{\isacharparenright}}\ {\isaliteral{5C3C6F723E}{\isasymor}}\ \isafun{afails}\ {\isaliteral{28}{\isacharparenleft}}G\ s{\isaliteral{29}{\isacharparenright}}{\isaliteral{29}{\isacharparenright}}} \label{eqn:merge-def}
\end{center}
\begin{figure}[tb]
\begin{tabular}{l c l@ {\hspace{24 pt}}c@ {\hspace{12 pt}}r}
\begin{minipage}[t]{0.12\textwidth}
\begin{isabelle}%
\isafun{doA}\isanewline
\ \ \ \ x\ {\isaliteral{5C3C6C6566746172726F773E}{\isasymleftarrow}}\ {\isaliteral{5C3C6C62726163653E}{\isasymlbrace}}P{\isaliteral{5C3C7262726163653E}{\isasymrbrace}}\ f{\isaliteral{3B}{\isacharsemicolon}}\isanewline
\ \ \ \ G\ x\isanewline
\isafun{odA}%
\end{isabelle}
\end{minipage}
&\isa{{\isaliteral{5C3C626F777469653E}{\isasymbowtie}}}&
\begin{minipage}[t]{0.12\textwidth}
\begin{isabelle}%
\isafun{doA}\isanewline
\ \ \ \ x\ {\isaliteral{5C3C6C6566746172726F773E}{\isasymleftarrow}}\ {\isaliteral{5C3C6C62726163653E}{\isasymlbrace}}Q{\isaliteral{5C3C7262726163653E}{\isasymrbrace}}\ f{\isaliteral{3B}{\isacharsemicolon}}\isanewline
\ \ \ \ G{\isaliteral{27}{\isacharprime}}\ x\isanewline
\isafun{odA}%
\end{isabelle}
\end{minipage}
&\isa{{\isaliteral{3D}{\isacharequal}}}&
\begin{minipage}[t]{0.2\textwidth}
\begin{isabelle}%
\isafun{doA}\isanewline
\ \ \ \ x\ {\isaliteral{5C3C6C6566746172726F773E}{\isasymleftarrow}}\ {\isaliteral{5C3C6C62726163653E}{\isasymlbrace}}P\ and\ Q{\isaliteral{5C3C7262726163653E}{\isasymrbrace}}\ f{\isaliteral{3B}{\isacharsemicolon}}\isanewline
\ \ \ \ G\ x\ {\isaliteral{5C3C626F777469653E}{\isasymbowtie}}\ G{\isaliteral{27}{\isacharprime}}\ x\isanewline
\isafun{odA}%
\end{isabelle}
\end{minipage}
\end{tabular}
\caption{Distributing an annotation merge across functions.\label{fig:merge-bind}}
\end{figure}
Note that only one of the functions \isa{F} is actually evaluated.
This is simply because merging is only sensible between
two annotations over the same function, and so
\isa{\isafun{dropA}\ F\ s\ {\isaliteral{3D}{\isacharequal}}\ \isafun{dropA}\ G\ s} is implicitly assumed.
We show in \autoref{fig:merge-bind} that this merge operation distributes
across function composition, so merging two annotations annotates
the individual operations. This can be repeatedly applied to show the result
in \autoref{fig:new-tcb-merge}. \isa{\isafun{new{\isaliteral{5F}{\isacharunderscore}}tcb}}
can then be shown to satisfy this merged annotation
using the following rule:
\begin{center}
\isa{\mbox{}\inferrule{\mbox{{\isaliteral{5C3C6C62726163653E}{\isasymlbrace}}P{\isaliteral{5C3C7262726163653E}{\isasymrbrace}}\ f\ {\isaliteral{5C3C737173756273657465713E}{\isasymsqsubseteq}}\ F}\\\ \mbox{{\isaliteral{5C3C6C62726163653E}{\isasymlbrace}}Q{\isaliteral{5C3C7262726163653E}{\isasymrbrace}}\ f\ {\isaliteral{5C3C737173756273657465713E}{\isasymsqsubseteq}}\ F{\isaliteral{27}{\isacharprime}}}}{\mbox{{\isaliteral{5C3C6C62726163653E}{\isasymlbrace}}P\ and\ Q{\isaliteral{5C3C7262726163653E}{\isasymrbrace}}\ f\ {\isaliteral{5C3C737173756273657465713E}{\isasymsqsubseteq}}\ F\ {\isaliteral{5C3C626F777469653E}{\isasymbowtie}}\ F{\isaliteral{27}{\isacharprime}}}}} \label{eqn:merge-naf-sub}
\end{center}
This states that a function satisfies the merge of two
annotations under the conjunction of their preconditions.

\subsection*{Automatic Annotations}

Significant amounts of a Hoare logic proof can be
automated by a vcg. At the core of any Hoare logic
vcg is a collection of rules that decompose a Hoare
triple into a collection of Hoare triples over
the individual steps of a function. In Isabelle, the weakest precondition
vcg \emph{wp} processes a function bottom-up, computing the
precondition necessary for the last operation to establish
the postcondition. This computed precondition then becomes
the postcondition for the next iteration, and so on.
This is accomplished by repeated applications of the split rule {\sc wp-split}.

To modify the vcg to work with annotators, we
introduce an analogous split rule which additionally 
saves the computed precondition as an annotation of \isa{f}.
\begin{equation}
\isa{\mbox{}\inferrule{\mbox{{\isaliteral{5C3C666F72616C6C3E}{\isasymforall}}x{\isaliteral{2E}{\isachardot}}\ {\isaliteral{5C3C6C62726163653E}{\isasymlbrace}}B\ x{\isaliteral{5C3C7262726163653E}{\isasymrbrace}}\ g\ x\ {\isaliteral{5C3C6C62726163653E}{\isasymlbrace}}C{\isaliteral{5C3C7262726163653E}{\isasymrbrace}}\ {\isaliteral{5C3C6C616E676C653E}{\isasymlangle}}G\ x{\isaliteral{5C3C72616E676C653E}{\isasymrangle}}}\\\ \mbox{{\isaliteral{5C3C6C62726163653E}{\isasymlbrace}}A{\isaliteral{5C3C7262726163653E}{\isasymrbrace}}\ f\ {\isaliteral{5C3C6C62726163653E}{\isasymlbrace}}B{\isaliteral{5C3C7262726163653E}{\isasymrbrace}}}}{\mbox{{\isaliteral{5C3C6C62726163653E}{\isasymlbrace}}A{\isaliteral{5C3C7262726163653E}{\isasymrbrace}}\ \isafun{do}\ \ \ \ \ x\ {\isaliteral{5C3C6C6566746172726F773E}{\isasymleftarrow}}\ f{\isaliteral{3B}{\isacharsemicolon}}\ \ \ \ \ g\ x\ \isafun{od}\ {\isaliteral{5C3C6C62726163653E}{\isasymlbrace}}C{\isaliteral{5C3C7262726163653E}{\isasymrbrace}}\ {\isaliteral{5C3C6C616E676C653E}{\isasymlangle}}\isafun{doA}\ \ \ \ \ x\ {\isaliteral{5C3C6C6566746172726F773E}{\isasymleftarrow}}\ {\isaliteral{5C3C6C62726163653E}{\isasymlbrace}}A{\isaliteral{5C3C7262726163653E}{\isasymrbrace}}\ f{\isaliteral{3B}{\isacharsemicolon}}\ \ \ \ \ G\ x\ \isafun{odA}{\isaliteral{5C3C72616E676C653E}{\isasymrangle}}}}} \label{eqn:valid-annotate-lifted-bind-atomized}
\end{equation}
Note that, after this rule is applied, 
only a standard Hoare triple is required to be shown of \isa{f}.
Repeated applications of this rule, combined with some additional
rules for control flow, result in a collection of Hoare triples as
proof obligations. When solved, the preconditions for these
Hoare triples are stored in the function annotations. 

To avoid ever explicitly stating the definition of an annotation
we can use Isabelle's \emph{schematic lemma} feature, which allows
terms in a lemma statement to be left unspecified and instantiated
during the proof. The bind rule given in~\eqref{eqn:valid-annotate-lifted-bind-atomized} will, after applied over the entire function,
instantiate such a term to the computed function annotation.%
\end{isamarkuptext}%
\isamarkuptrue%
\begin{isamarkuptext}%
For example, we rephrase an existing Hoare triple proof as
an annotator, leaving the annotation as a schematic variable.
\begin{center}
\isa{{\isaliteral{5C3C6C62726163653E}{\isasymlbrace}}\isafun{even}\ i{\isaliteral{5C3C7262726163653E}{\isasymrbrace}}\ \isafun{double{\isaliteral{5F}{\isacharunderscore}}plus}\ i\ {\isaliteral{5C3C6C62726163653E}{\isasymlbrace}}{\isaliteral{5C3C6C616D6264613E}{\isasymlambda}}{\isaliteral{5F}{\isacharunderscore}}{\isaliteral{2E}{\isachardot}}\ \isafun{even}\ i{\isaliteral{5C3C7262726163653E}{\isasymrbrace}}\ {\isaliteral{5C3C6C616E676C653E}{\isasymlangle}}{\isaliteral{3F}{\isacharquery}}L{\isaliteral{5C3C72616E676C653E}{\isasymrangle}}}
\end{center}
After the proof is finished, the schematic has been instantiated
to a function annotation.
\begin{center}
\isa{{\isaliteral{5C3C6C62726163653E}{\isasymlbrace}}\isafun{even}\ i{\isaliteral{5C3C7262726163653E}{\isasymrbrace}}\ \isafun{double{\isaliteral{5F}{\isacharunderscore}}plus}\ i\ {\isaliteral{5C3C6C62726163653E}{\isasymlbrace}}{\isaliteral{5C3C6C616D6264613E}{\isasymlambda}}{\isaliteral{5F}{\isacharunderscore}}{\isaliteral{2E}{\isachardot}}\ \isafun{even}\ i{\isaliteral{5C3C7262726163653E}{\isasymrbrace}}\ {\isaliteral{5C3C6C616E676C653E}{\isasymlangle}}\isafun{doA}\ \ \ \ \ {\isaliteral{5C3C6C62726163653E}{\isasymlbrace}}\isafun{even}\ i{\isaliteral{5C3C7262726163653E}{\isasymrbrace}}\ i{\isaliteral{2B}{\isacharplus}}{\isaliteral{2B}{\isacharplus}}{\isaliteral{3B}{\isacharsemicolon}}\ \ \ \ \ {\isaliteral{5C3C6C62726163653E}{\isasymlbrace}}\isafun{odd}\ i{\isaliteral{5C3C7262726163653E}{\isasymrbrace}}\ i{\isaliteral{2B}{\isacharplus}}{\isaliteral{2B}{\isacharplus}}\ \isafun{odA}{\isaliteral{5C3C72616E676C653E}{\isasymrangle}}}
\end{center}

\subsection*{Complex Annotations}
Function annotations defined in this way are simply extensions
of existing functions, which track an additional boolean across all
possible nondeterministic branches. Due to this construction, they
are necessarily as expressive as the underlying monadic formalization. In particular
one can annotate a recursive function (creating a recursive annotation) or a map over
an inductive datatype. This would serve to simplify induction hypotheses while reasoning
about these functions as necessary correctness invariants could be assumed across all
iterations of the function. These annotations could still be collected automatically
from existing proofs, although the process would be slightly more involved. The use
of function annotations in this context seems promising, but has not been fully explored
and has been left for future work.

\section{Case Study: seL4}

Interesting properties cannot always be expressed in Hoare logic.
In the ongoing proof of confidentiality for the seL4 
microkernel~\cite{Murray_MBGK_12},
a proof calculus was used to formalize an upper bound on the
information that a function can read. This involves reasoning
about multiple executions of a function, which cannot be
easily expressed in Hoare logic. Function annotations can
still be used when proving such a property by explicitly turning
annotations into assertions, effectively converting back
into a standard function while retaining some information
from the annotations. The standard confidentiality 
calculus~\cite{Murray_MBGK_12} can then be applied and make use of the assertions.

The confidentiality proof for seL4 builds on previous
invariant proofs in addition to proofs of integrity and
authority confinement~\cite{Sewell_WGMAK}. In general, for a function to
be confidentiality-preserving it must be well-behaved,
which requires reasoning about it in the presence of invariants.
These invariants are then required as preconditions for
confidentiality and we must therefore demonstrate that
 certain invariants are preserved at intermediate points of
a function. In the majority of cases there are sufficient
existing lemmas to easily re-play this reasoning,
however some functions have behaviours which are difficult
to characterize. Such a function is \isa{\isafun{invoke{\isaliteral{5F}{\isacharunderscore}}untyped}},
which changes the type of kernel objects in a region of memory. The proof that \isa{\isafun{invoke{\isaliteral{5F}{\isacharunderscore}}untyped}} preserves
the invariants is 300 lines of Isabelle proof script. Some manual
effort was involved to re-phrase the top-level precondition
(approximately 5 lines of Isabelle) in order to produce a function annotation from this proof. Additionally,
 the generated annotation
needed to be manually modified in order to remove
extraneous properties. Ultimately a single line was added to the script
with 40 lines of manual annotation modifications. When this annotation
was applied to the proofs of integrity and authority confinement, the
number of lines of proof script was reduced from 192 to 56. This
produced another annotation which, when applied to the proof of
confidentiality, reduced the lines of proof script from 215 to to 39. 
In this case it is clear that most of the logic in these proofs was simply re-proving the preservation of the invariants.
Some manual effort is still required to effectively use
these annotations, so they are currently in an experimental branch
of the seL4 proof. There has been some development in adding
more automation to this process, but it is still ongoing.%
\end{isamarkuptext}%
\isamarkuptrue%
\isadelimtheory
\endisadelimtheory
\isatagtheory
\endisatagtheory
{\isafoldtheory}%
\isadelimtheory
\endisadelimtheory
\end{isabellebody}%

\section{Related and Future Work}
Similar to our use of nondeterministic state
monads~\cite{Cock_KS_08}, Sprenger et al.~\cite{Sprenger_B_07} formalize Hoare logic over state monads
in the verification of a security protocol. They similarly
use a shallow embedding to exploit the significant amount
of proof automation already present in Isabelle/HOL.

Swierstra~\cite{Swierstra_09} encodes the correctness of a specification in
its type, creating a \emph{strong specification}. The
construction of the specification itself guarantees
correctness, with respect to a given property. In our
logic, the relationship between a function and its annotations
is analogous to the relationship between a specification and
its strong specification.

Function annotations are merely one approach to solving the general
problem of reusing results about the intermediate states of a function.
Alternatively, assertions could be explicitly provided as part of
the specification, as done by Mossakowski et al.~\cite{Mossakowski_TSLGS_08} in their formalization of Hoare logic on monads. Although
this approach is more general, it introduces barriers
to splitting up the same proof among multiple persons~\cite{Cock_KS_08}.
Additionally, extending these assertions to include further properties
may incur significant proof maintenance overhead. In contrast,
 our formalization produces annotations as an orthogonal 
artifact to the specification itself.

Another approach would be to have intermediate states
 explicitly labelled and assertions made against those labels.
This would enable the ad-hoc collection of facts about intermediate
states, as is done by using function annotations. Rather than 
constructing several annotated versions of a function, the user
would instead be creating an assertion cache for the intermediate states.
Such an approach would allow for more dynamic use of assertions, as
the user would not have to decide at the 
start of a proof which annotation sets he wished to use.
The cost, however, comes in the initial overhead of
defining these labels. The most straightforward approach would
be to modify the monad formalization to label intermediate states, which
would incur an initial overhead for any existing proofs. In the case
of a large proof effort like seL4 this could potentially affect hundreds of thousands of 
lines of proof script.
The advantage of function annotations is their low overhead, as they do not require
any modification to the underlying formalization.

Function annotations as presented here have been shown
to work well in practice, however some manual
mechanical effort is still required to make use of them.
In future work we plan to make function annotations
more opaque to the end user. Ideally, rather than being explicitly
exported through schematic lemmas, they could implicitly be generated as part
of standard Hoare logic proofs. During further proofs
over the same function, annotations could then implicitly be used by the vcg, or asserted if explicit reasoning needs to be done against them.
Additionally, properties could be tagged as ``annotation-worthy"
and will otherwise not be pushed into automatic annotation generation.
With these improvements, the user would not need to be aware that 
they are using function annotations, but could transparently
exploit previously established reasoning with a collection of tactics and through the vcg.

\section{Conclusion}
In this paper we have presented a function annotation logic
which can be used to prove properties about intermediate
points of functions. It is designed to generate 
annotations from existing Hoare logic proofs with minimal 
modifications to these proofs, allowing the reasoning
within those proofs to be easily re-used.
Annotations can be used to prove additional,
stronger annotations and unrelated annotations can be merged
together as a conjunction of their individual assertions. They
are useful when a function is too complex for a vcg alone and re-using
the reasoning of previous proofs can save significant effort.
The seL4 proofs for two properties of a single function were reduced from 407 lines of proof script to 95 by using function annotations. This resulted in more comprehensible
proofs that better captured the logic of the properties being shown. By extending the use of function annotations to additional
functions of similar complexity, we expect to see comparable
improvements to existing and future proofs.

\section*{Acknowledgements}
Thanks to Toby Murray and Gerwin Klein for valuable
feedback on earlier drafts of this paper.

\bibliographystyle{eptcs}
\bibliography{references}
\end{document}